\documentclass[twocolumn,10pt,amsmath,amssymb,aps,pra,superscriptaddress]{revtex4-1}
\usepackage{textcomp}
\usepackage{graphicx}
\usepackage{longtable}
\usepackage{color}
\usepackage{dcolumn}
\usepackage{multirow}
\usepackage{epstopdf}
\usepackage{bm}
\usepackage{hyperref}
\usepackage{nicefrac}
\newcolumntype{d}[1]{D{.}{.}{#1}}

\newcommand{\ninej}[9]
{
  \left\{ \begin{array}{ccc} #1 & #2 & #3 \\
    #4 & #5 & #6 \\ #7 & #8 & #9 \end{array} \right\}
}
\newcommand{\sixj}[6]
{
  \left\{ \begin{array}{ccc} #1 & #2 & #3 \\
    #4 & #5 & #6 \end{array} \right\}
}

\begin{document}

\title{Purely long-range polar molecules composed of identical\\
 lanthanide atoms}

\author{Hui Li}
\affiliation{Laboratoire Aim\'{e} Cotton, CNRS, Universit\'{e} Paris-Sud, ENS Paris-Saclay,  Universit\'{e}  Paris-Saclay, 91405 Orsay, France}
\affiliation{Department of Physics, Temple University, Philadelphia, PA 19122, USA}

\author{Goulven Qu{\'e}m{\'e}ner}
\affiliation{Laboratoire Aim\'{e} Cotton, CNRS, Universit\'{e} Paris-Sud, ENS Paris-Saclay,  Universit\'{e}  Paris-Saclay, 91405 Orsay, France}

\author{Jean-Fran{\c c}ois Wyart}
\affiliation{Laboratoire Aim\'{e} Cotton, CNRS, Universit\'{e} Paris-Sud, ENS Paris-Saclay,  Universit\'{e}  Paris-Saclay, 91405 Orsay, France}
\affiliation{LERMA, Observatoire de Paris-Meudon, PSL Research University, Sorbonne Universit{\'e}s, UPMC Univ.~Paris 6, CNRS UMR8112, 92195 Meudon, France}

\author{Olivier Dulieu}
\affiliation{Laboratoire Aim\'{e} Cotton, CNRS, Universit\'{e} Paris-Sud, ENS Paris-Saclay,  Universit\'{e}  Paris-Saclay, 91405 Orsay, France}

\author{Maxence Lepers}
\email{maxence.lepers@u-bourgogne.fr}
\affiliation{Laboratoire Aim\'{e} Cotton, CNRS, Universit\'{e} Paris-Sud, ENS Paris-Saclay,  Universit\'{e}  Paris-Saclay, 91405 Orsay, France}
\affiliation{Laboratoire Interdisciplinaire Carnot de Bourgogne, CNRS, Universit\'e de Bourgogne Franche-Comt\'e, 21078 Dijon, France}


\begin{abstract}
Doubly polar molecules, possessing an electric dipole moment and a magnetic dipole moment, can strongly couple to both an external electric field and a magnetic field, providing unique opportunities to exert full control of the system quantum state at ultracold temperatures. We propose a method for creating a purely long-range doubly polar homonuclear molecule from a pair of strongly magnetic lanthanide atoms, one atom being in its ground level and the other in a superposition of quasi-degenerate opposite-parity excited levels [Phys.~Rev.~Lett.~\textbf{121}, 063201 (2018)]. The electric dipole moment is induced by coupling the excited levels with an external electric field. We derive the general expression of the long-range, Stark, and Zeeman interaction energies in the properly symmetrized and fully-coupled basis describing the diatomic complex. Taking the example of holmium, our calculations predict shallow long-range wells in the potential energy curves that may support vibrational levels accessible by direct photoassociation from pairs of ground-level atoms.
\end{abstract}


\maketitle

\section{Introduction}

A unique feature of ultracold quantum gases is the tunability of the interaction strength between particles with external fields. Polar molecules, with numerous degrees of freedom and strong anisotropic interactions, represent an ideal platform  for applications such as the realization of new quantum many-body systems, precise tests of fundamental theories, controlled quantum chemistry, quantum simulation and quantum information \cite{demille2002, dulieu2009, carr2009, baranov2012, quemener2012, tscherbul2015, moses2016, gadway2016, bohn2017, wolf2017}. To this end, various heteronuclear bialkali polar molecules have been produced in their ground state over the last decade \cite{ni2008, deiglmayr2008a, takekoshi2012, wu2012, molony2014, park2015, guo2016}. Due to the absence an electronic magnetic dipole moment in their singlet ground state, the weak magnetic moments originating from nuclear spins, and weak non-linear Stark effect of their rovibrational ground level, they can not be easily manipulated by external fields.

Presently ultracold molecules with both electric and magnetic dipole moments are receiving burgeoning interest because of the greater possibilities for trapping and manipulation \cite{tscherbul2006a, micheli2006, perez2010, brue2012, pasquiou2013, roy2016, reens2017}. The simplest kind of such molecules are diatomics with an electronic ground state of $^2\Sigma$ and $^3\Sigma$ symmetry. The most promising $^2\Sigma$ species consists in pairing alkali-metals with divalent atoms such as the alkaline earth or ytterbium atoms \cite{tassy2010, zuchowski2010a, hara2011, ivanov2011, brue2013, pasquiou2013, roy2016, gerschmann2017, guttridge2017}, the prime example being RbSr which has been predicted to have a permanent electric dipole moment of 1.4-1.5 debye (D) \cite{guerout2010, zuchowski2010a}. Recent highlights in this candidate include a quite advanced ongoing experiment in which the magnetic Feshbach resonances have been observed in the corresponding atomic mixtures \cite{barbe2018}, and a quite promising theoretical modeling \cite{devolder2018}. Example of $^3\Sigma$ species can be found with the energetically-lowest a$^3\Sigma$ state of heteronuclear bialkali dimers, for instance LiNa which has been successfully created in the ultracold regime \cite{rvachov2017}. Metastable LiNa possesses a magnetic dipole moment of 2$\mu_B$ as well as an electric dipole moment of 0.2 D \cite{aymar2005, tomza2013b}, where $\mu_B$ is Bohr's magneton.

Meanwhile, highly magnetic atoms, such as chromium and open-shell lanthanides, have also brought new perspectives in the field of ultracold quantum gases and provided the opportunity to explore the behavior of long-range interacting polar systems beyond previously accessible regimes \cite{sukachev2010, lu2011b, aikawa2012, miao2014, baier2016, kadau2016, dreon2017}. The diatomic molecules containing chromium, such as CrRb \cite{pavlovic2010}, CrSr and CrYb \cite{tomza2013}, or containing lanthanide atoms, such as Eu-alkali metal dimers \cite{tomza2014, zaremba2018} and ErLi \cite{gonzalez2015}, have been theoretically proposed as candidates with both large magnetic and electric dipole moments. Experimentally, pairs of highly magnetic atoms Er$_2$, with a strong magnetic dipole moment up to 12$\mu_B$ \cite{frisch2015} have been produced in a weakly bound level, and photoassociation into spin-polarized Cr$_2$ dimers has also been demonstrated in Ref.~\cite{ruhrig2016}. More recently, magnetoassociation into ultracold Eu$_2$ dimers were also theoretically investigated \cite{zaremba2018}. Besides, the realization of ultracold mixtures of Dy and K atoms \cite{ravensbergen2018} and Dy and Er atoms \cite{ilzhofer2018} opens up new possibilities for forming ultracold molecules in nontrivial electronic states.

In a recent work, we have demonstrated the possibility to induce a strong electric dipole moment, up to 0.22 D, on dysprosium atoms \cite{lepers2018a}, in addition to a large magnetic dipole moment of 13$\mu_B$, by preparing the atoms in a superposition of nearly-degenerate opposite-parity excited levels, which are mixed with an external electric field. In the present article, we extend our work to the production of doubly-polar homonuclear molecules by binding the excited atom to a ground-level one. Due to the current inability to calculate potential energy curves between pairs of open-shell lanthanides at small internuclear distances, we explore the possibility to form purely long-range molecules as demonstrated with pairs of alkali metals and close-shell atoms \cite{movre1977, stwalley1978, leonard2003, enomoto2008}. Here we choose holmium, as it possesses a pair of nearly degenerate opposite-parity levels, accessible from the ground level by a strong one-photon transition, which opens the possibility to form the long-range molecules by direct photoassociation \cite{jones2006}.

In this article, we characterize the long-range interactions between two identical atoms, one being in the ground level and the other being in a superposition of nearly degenerate opposite-parity excited levels that are coupled by an external electric field. We present the formalism to calculate the potential energy for interactions between arbitrary atomic multipoles, as well as Stark and Zeeman interactions, in the fully-coupled and properly symmetrized diatomic basis including hyperfine structure. The potential-energy curves that we compute present shallow long-range wells with a few vibrational levels, 
strong magnetic moments and non-zero induced electric dipole moments, even though the considered molecule is homonuclear \cite{greene2000, li2011}.
 We also find numerous repulsive curves that may be used for optical shielding of collisions between ground-level holmium atoms \cite{suominen1995, quemener2016, karman2018, lassabliere2018, orban2019}.

The structure of this article is as follows. In section \ref{sec:thf} we describe the theoretical formalism for two interacting atoms in the presence of external electric and magnetic fields, including a general presentation (Subsection \ref{sub:gene}), symmetrization of basis functions (Subsection \ref{sub:basis}) and matrix elements of the Hamiltonian (Subsection \ref{sub:intp}). Then in section \ref{sec:rd}, we apply our formalism to characterize the interactions between two holmium atoms, by calculating potential-energy curves, vibrational levels and induced electric dipole moments. Section \ref{sec:conclu} contains concluding remarks.

\section{Theory}
\label{sec:thf}

\subsection{General form of the Hamiltonian}
\label{sub:gene}

We consider two identical atoms with nuclear spin $I$ interacting with each other. One atom is in the ground level $|g \rangle$ with electronic and total angular momenta $J_g$ and $F_g$, whereas the other is excited in a superposition of two quasi-degenerate opposite-parity levels, labeled $|a\rangle$ ($|b\rangle$) with energy $E_{a}$ ($E_{b}$), electronic and total angular momenta $J_{a}$ ($J_{b}$) and $F_{a}$ ($F_{b}$). We assume that $|a\rangle$ has the same electronic parity as $|g \rangle$. 
For open-$4f$-shell lanthanide atoms, the electronic angular momentum is large, \textit{e.g.}~$J_{g}=J_{a}=15/2$ and $J_{b}=17/2$ for holmium.

The model Hamiltonian for a system of two interacting atoms at large distances with reduced mass $\mu$, internuclear separation $R$ and relative angular momentum $\hat{L}$, can be expressed as, 
\begin{equation}
  \hat{H} = -\frac{1}{2\mu R}\frac{\partial^2 }{\partial R^2}R
    + \frac{\hat{\bf{L}}^{2}}{2\mu R^{2}} + \sum_{i=1}^{2} \hat{H}_{i}
    + \hat{V}_{LR}\left(R\right) \,.
  \label{eq:hamilt}
\end{equation}
The first two terms are the radial and angular parts of the kinetic-energy operator; the last term is the long-range potential energy, and $\hat{H}_i$ the Hamiltonian of individual atom $i$. In the presence of external electric and magnetic fields, $\hat{H}_i$ can be written as
\begin{equation}
  \hat{H}_i = \hat{H}_{hf}(i) + \hat{H}_{S}(i) + \hat{H}_{Z}(i)
\end{equation}
where the first term is the field-free atomic Hamiltonian including hyperfine interactions, the second and third terms are Stark and Zeeman interactions. In the coupled atomic basis $|J_i I F_i M_{F_i}\rangle$, the hyperfine interactions are diagonal with energies \cite{drake2006},
\begin{eqnarray}
  E_{hf}(i) & = & \frac{1}{2}A_i C_i 
  \nonumber \\
   & + & B_i\frac{3/4 \times C_i(C_i+1)-I(I+1)J_i(J_i+1)}{2I(2I-1)J_i(2J_i-1)}
\end{eqnarray}
where $C_i=F_i(F_i+1)-J_i(J_i+1)-I(I+1)$, $A_i$ and $B_i$ are the hyperfine structure constants. The matrix elements of $\hat{H}_{S}(i)$ and $\hat{H}_{Z}(i)$ will be given below.

\begin{table*}
	\caption{Block structure of the potential energy matrix, comprising Stark, Zeeman and long-range dipolar and quadrupolar interactions between two identical atoms, one in the ground level $|g\rangle$ and the other in a superposition of opposite-parity excited levels $|a\rangle$ and $|b\rangle$. The notations have the following meaning: $\hat{H}_{S}$ and $\hat{H}_{Z}$ are the Stark and Zeeman interactions, $V_{dd}$, $V_{dq}$, $V_{qd}$ and $V_{qq}$ are the electric dipole-dipole, dipole-quadrupole, quadrupole-dipole and quadrupole-quadrupole interactions, $V_{\mu\mu}$ is the magnetic dipole-dipole interaction; the superscripts {}``dir'' and {}``res'' correspond to direct and resonant interactions respectively.
		\label{tab:uncp}}
	\centering
\begin{ruledtabular}
		\begin{tabular}{ccccc}
	                      & $|g \rangle |a \rangle$ & $|g \rangle |b \rangle$ & $|a \rangle |g \rangle$ & $|b \rangle |g \rangle$ \\
\hline
  $\langle g| \langle a|$ & $\hat{H}_{Z} + \hat{V}_{qq}^{dir}(R) + \hat{V}_{\mu\mu}^{dir}(R)$ & $\hat{H}_{S}(2) + \hat{V}_{qd}^{dir}(R)$ & $\hat{V}_{qq}^{res}(R)$ & $\hat{V}_{dq}^{res}(R)$  \\
  $\langle g| \langle b|$ & $\hat{H}_{S}(2) + \hat{V}_{qd}^{dir}(R)$ & $\hat{H}_{Z} + \hat{V}_{qq}^{dir}(R) + \hat{V}_{\mu\mu}^{dir}(R)$ & $\hat{V}_{qd}^{res}(R)$ & $\hat{V}_{dd}^{res}(R)$  \\
  $\langle a| \langle g|$ & $\hat{V}_{qq}^{res}(R)$ & $\hat{V}_{qd}^{res}(R)$ & $\hat{H}_{Z} + \hat{V}_{qq}^{dir}(R) + \hat{V}_{\mu\mu}^{dir}(R)$ & $\hat{H}_{S}(1) + \hat{V}_{dq}^{dir}(R)$  \\
  $\langle b| \langle g|$ & $\hat{V}_{dq}^{res}(R)$ & $\hat{V}_{dd}^{res}(R)$ & $\hat{H}_{S}(1) + \hat{V}_{dq}^{dir}(R)$ & $\hat{H}_{Z} + \hat{V}_{qq}^{dir}(R) + \hat{V}_{\mu\mu}^{dir}(R)$  \\
		\end{tabular}
	\end{ruledtabular}
\end{table*}

Having a large electronic angular momentum, the levels $|g\rangle$, $|a\rangle$ and $|b\rangle$ possess permanent magnetic dipole and electric quadrupole moments. Moreover, $|a\rangle$ and $|b\rangle$ are coupled by electric-dipole transition, due to their opposite parity. The electric and magnetic dipole moments give respectively rise to Stark and Zeeman shifts. The dipole and quadrupole moments also give rise to direct interactions in the multipolar expansion.

Because the two atoms are identical but in different quantum levels, they also interact via resonant terms of the multipolar expansion \cite{lepers2018}. (i) The levels $|g\rangle$ and $|b\rangle$, of opposite parities, show a resonant electric dipole-dipole interaction, scaling as $R^{-3}$; (ii) the levels $|g\rangle$ and $|a\rangle$, of identical parity, show a resonant electric quadrupole-quadrupole interaction, scaling as $R^{-5}$; and (iii) an electric dipole, coupling $|g\rangle$ and $|b\rangle$, and an electric quadrupole, coupling $|g\rangle$ and $|a\rangle$, resonantly interact with an energy scaling as $R^{-4}$.

The direct and resonant atom-atom interactions are schematically summarized in Table \ref{tab:uncp}, as well as the field-atom interactions. The basis functions are divided in four blocks $|g \rangle |a \rangle$, $|g \rangle |b \rangle$, $|a \rangle |g \rangle$, $|b \rangle |g \rangle$, where $|g\rangle$, $|a\rangle$, $|b\rangle$ stands for all the quantum numbers of the corresponding levels. The other quantum numbers, \textit{e.g.}~the partial wave $L$, are not shown. The direct and atom-field interactions are located in the top-left and bottom-right parts of the table, while resonant interactions are located in the top-right and bottom-left parts.

\subsection{Basis sets and symmetries}
\label{sub:basis}

Each atom $i$ ($i=1, 2$) is described by its total electronic $\hat{\bm{J}_i}$ and nuclear spin $\hat{\bm{I}}$ (identical for the two atoms), which combine to form the total atomic angular momentum $\hat{\bm{F}_i} = \hat{\bm{J}_i} + \hat{\bm{I}}$. The associated quantum numbers are $J_i$, $I$ and $F_i$. The projections of the angular momenta are considered along the $z$ axis of space-fixed coordinate system; the associated quantum numbers are $M_{J_i}$, $M_{I_i}$ and $M_{F_i}$. The electronic parity $p_i = \pm 1$ under inversion of electronic coordinates is identical for levels $|g\rangle$ and $|a\rangle$ and opposite for $|g\rangle$ and $|b\rangle$. Finally, the angular momentum $\hat{\bm{L}}$ accounts for the rotation of the internuclear axis in the space-fixed frame. Its magnitude is associated with the partial wave $L$ and its $z$-projection with $M_L$. Therefore we obtain the uncoupled basis $|\beta_1 J_1 I F_1 M_{F_1}\rangle |\beta_2 J_2 I F_2 M_{F_2}\rangle |L M_L\rangle \equiv |\beta_1 \beta_2 J_1 I F_1 M_{F_1} J_2 I F_2 M_{F_2} L M_L\rangle$, where $\beta_1$ and $\beta_2$ gather all the other quantum numbers of atoms 1 and 2 (parities are not explicitly written).

For convenience, we perform the present calculations in the fully-coupled basis. Indeed for a diatomic system in the long-range region, the electronic angular momentum of each atom $\hat{\bm{J}}_i$ is more strongly coupled to the nuclear spin $\hat{\bm{I}}$ than to the internuclear axis \cite{enomoto2008}. We thus introduce the coupled angular momentum $\hat{\bm{F}}_{12} = \hat{\bm{F}}_1 + \hat{\bm{F}}_2$, itself composed with $\hat{\bm{L}}$ to give the total angular momentum of the complex $\hat{\bm{F}} = \hat{\bm{F}}_{12} + \hat{\bm{L}}$. The resulting fully-coupled basis functions $|\beta_1 \beta_2 J_1 I_1 F_1 J_2 I_2 F_2 F_{12} L F M_F \rangle $ are related to the uncoupled by Equation \eqref{eq:cplbas} of Appendix \ref{sec:cplbas}. In absence of external field, the total angular momentum $F$ is a good quantum number; here the field amplitude are low enough, so that the different $F$ values are weakly coupled. For 	fields parallel the $z$ axis, the total angular-momentum projection $M_F = M_{F_1} + M_{F_2} + M_L = M_{F_{12}} + M_L$ is conserved. Among the basis functions, one can distinguish between even and odd ones with respect to the inversion of all the electronic and nuclear coordinates. Namely a given function has a total parity of $p_1p_2(-1)^{L}$, which is not a strictly good quantum number because of the electric field; still, even and odd functions are not strongly coupled in the range of field amplitudes considered here. Finally for $M_F=0$, one has even and odd basis functions with respect to the reflection about the space-fixed $xz$ plane, depending on whether $p_1p_2(-1)^{L+F}$ is equal to $+1$ or $-1$ (see Appendix \ref{sec:cplbas}).

For systems of identical particles, the permutation symmetry must be taken into account \cite{tscherbul2009}. We build the properly symmetrized fully-coupled basis for the two identical atoms (see detailed discussion in Appendix \ref{sec:cplbas}),
\begin{widetext}
\begin{align}
  |\beta_1\beta_2J_1IF_1J_2IF_2F_{12}LFM_F;\eta \rangle =
    & \frac{1} {\sqrt{2(1+\delta_{\beta_1\beta_2} \delta_{J_1J_2} \delta_{F_1F_2})}} 
      \left \{|\beta_1\beta_2J_1IF_1J_2IF_2F_{12}LFM_F\rangle \right.
  \nonumber \\
    & \left. + \eta (-1)^{F_1+F_2-F_{12}+L}
    |\beta_2\beta_1J_2IF_2J_1IF_1F_{12}LFM_F\rangle \right \}
  \label{eq:sym-basis}
\end{align}
\end{widetext}
The symmetry of the basis functions with respect to the permutation of the identical atoms is given by index $\eta$: for bosonic isotopes, only the value $\eta=+1$ is allowed, while $\eta=-1$ for the fermionic ones.

In the following we will first construct the Hamiltonian in the fully coupled basis, then we will transform the Hamiltonian to the symmetrized basis by using the method from Ref.~\cite{green1975, alexander1977, quemener2011b}.

\subsection{Matrix element of the Hamiltonian in the fully coupled basis}
\label{sub:intp}

In this subsection, matrix elements will be given in the unsymmetrized fully-coupled basis. Going to the symmetrized one requires to apply Eq.~\eqref{eq:sym-basis} in the bras and the kets of the matrix elements.

\subsubsection{Atomic multipole moments}

The Stark, Zeeman, and long-range Hamiltonians are functions of 
the electric and magnetic multipole-moment operators $\hat{Q}_{\ell_i m_i}$ and $\hat{\mu}_{\ell_i m_i}$ of atoms $i=1$ and 2.
 Since they are irreducible tensors of rank $\ell_i$ and component $m_i$, their matrix elements satisfy the Wigner-Eckart theorem \cite{varshalovich1988}
\begin{align}
  & \langle \beta'_i J'_i I F'_i M'_{F_i} | \hat{Q}_{\ell_i m_i} 
    | \beta_i J_i I F_i M_{F_i} \rangle \notag \\
  & = \frac{C_{F_iM_{F_i}\ell_im_i}^{F'_iM'_{F_i}}} {\sqrt{2F'_i+1}}
    \langle \beta'_i J'_i I F'_i \| \hat{Q}_{\ell_i}
    \| \beta_i J_i I F_i \rangle
  \label{eq:qlm-we-1}
\end{align}
where $C_{a\alpha b\beta}^{c\gamma} = \langle a\alpha b\beta |abc\gamma \rangle$ is a Clebsch-Gordan coefficient \cite{varshalovich1988}, $\langle \beta'_i J'_i I F'_i \| \hat{Q}_{\ell_i} \| \beta_i J_i I F_i \rangle$ is the reduced matrix element (and similarly for $\hat{\mu}_{\ell_i}$). Assuming that the multipole moments are purely electronic operators, \textit{i.e.}~leaving the nuclear spin unchanged, the reduced matrix element can be expressed as
\begin{align}
  & \left \langle \beta'_i J'_i I F'_i \left \| \hat{Q}_{\ell_i}
    \right \| \beta_i J_i I F_i \right \rangle \notag \\
  & \quad = (-1)^{I+F_i+\ell_i+J'_i} \sqrt{(2F_i+1)(2F'_i+1)} \notag \\
  & \quad \times \sixj{J_i}{I}{F_i}{F'_i}{\ell_i}{J'_i}
    \left \langle \beta'_i J'_i \left \| \hat{Q}_{\ell_i}
    \right \| \beta_i J_i \right \rangle ,
  \label{eq:qlm-we-2}
\end{align}
where the quantity between curly brackets is a Wigner 6-j symbol \cite{varshalovich1988}. In this work, we deal with the electric dipole $\hat{Q}_1$ and quadrupole moments $\hat{Q}_2$, and the magnetic dipole moment $\hat{\mu}_1$ such that \cite{varshalovich1988}
\begin{align}
  \left \langle \beta_i J_i \left \| \hat{\mu}_1
    \right \| \beta_i J_i \right \rangle & 
    = -\mu_B g_i \langle \beta_i J_i \| \hat{\bm{J}_i}
    \| \beta_i J_i \rangle \notag \\
  & = -\mu_B g_{i} \sqrt{J_i(J_i+1)(2J_i+1)}
  \label{eq:magmom}
\end{align}
where $\mu_B$ is Bohr's magneton, and $g_{i} \equiv g_{J_i}$ is the electronic Land{\'e} $g$-factor of the level $|i\rangle$.

\subsubsection{Stark Hamiltonian}

We consider a homogeneous electric field $\bm{E} = \mathcal{E} \bm{u}_z$ oriented along the $z$ direction. At the first-order of perturbation the Stark effect operator can be written as $\hat{H}_S(i) = -\mathcal{E} \hat{Q}_{10}(i)$, where $\hat{Q}_{10}$ is the $z$-component of the dipole moment operator. Now assuming atom 1 in the ground level $|g\rangle$, and atom 2 in a superposition of excited levels $|a\rangle$ and $|b\rangle$, the matrix element of the Stark Hamiltonian $\hat{H}_S(2)$ in the uncoupled basis can be expressed as,
%
\begin{widetext}
\begin{align}
  & \left \langle \beta'_1 \beta'_2 J'_1 I F'_1 M'_{F_1} J'_2 I F'_2 M'_{F_2} L' M'_L
    \left | \hat{H}_S (2) \right | 
    \beta_1 \beta_2 J_1 I F_1 M_{F_1} J_2 I F_2 M_{F_2} L M_L \right \rangle
  \nonumber \\	
  & = -\mathcal{E} \delta_{\beta'_1 \beta_1} \delta_{J'_1 J_1} \delta_{F'_1 F_1} 
    \delta_{M'_{F_1} M_{F_1}} \delta_{L' L} \delta_{M'_L M_L} 
    \left \langle \beta'_2 J'_2 I F'_2 M'_{F_2} \left | \hat{Q}_{10} (2) 
    \right | \beta_2 J_2 I F_2 M_{F_2} \right \rangle 
  \nonumber \\
  & = -\mathcal{E} \delta_{\beta'_1 \beta_1} \delta_{J'_1 J_1} \delta_{F'_1 F_1}
    \delta_{M'_{F_1} M_{F_1}} \delta_{L' L} \delta_{M'_L M_L}
    \frac{C_{F_2 M_{F_2}10}^{F'_2 M'_{F_2}}}{\sqrt{2F'_2 +1}}
    \times \left \langle \beta'_2 J'_2 I F'_2 \left \| \hat{Q}_1 (2) 
    \right \| \beta_2 J_2 I F_2 \right \rangle 
\end{align}
The Clebsch-Gordan coefficient $C_{F_2 M_{F_2}10}^{F'_2 M'_{F_2}}$, which imposes $M'_{F_2} = M_{F_2}$, ensures the conservation of the atomic angular momentum projection along $z$. Using the sums involving two and three Clebsch-Gordan coefficients (see Appendix \ref{sec:cg}) and the reduced matrix elements of Eqs.~\eqref{eq:qlm-we-1} and \eqref{eq:qlm-we-2}, we can derive the matrix elements of $\hat{H_S}(1)$ and $\hat{H_S}(2)$ in the fully-coupled basis
\begin{align}
  & \left \langle \beta'_1 \beta'_2 J'_1 I F'_1 J'_2 I F'_2 F'_{12} L' F' M'_{F }
    \left | \hat{H}_{S}(2) \right | \beta_1 \beta_2 J_1 I F_1 J_2 I F_2 F_{12} L F M_F 
    \right \rangle \nonumber \\
  & = \mathcal{E} \delta_{\beta'_1 \beta_1} \delta_{J'_1 J_1} \delta_{F'_1 F_1} 
    \delta_{L' L} \delta_{M'_{F} M_F} 
    (-1)^{I+J'_2-F_1+L+F}
    \sqrt{(2F'_2 +1)(2F_2 +1)(2F'_{12} +1)(2F_{12} +1)(2F+1)} \nonumber \\ 
  & \times \sixj{J_2}{I}{F_2}{F'_2}{1}{J'_2} 
    \sixj{F_2}{F_1}{F_{12}}{F'_{12}}{1}{F'_2} \sixj{F_{12}}{L}{F}{F'}{1}{F'_{12}} 
    C_{FM_F 10}^{F' M_F} \left \langle \beta'_2 J'_2 \left \| \hat{Q}_{1}(2) 
    \right \|\beta_2 J_2 \right \rangle ,
\label{eq:stark-fcp-2} 
\end{align}	
and
\begin{align}
  &	\left \langle \beta'_1 \beta'_2 J'_1 I F'_1 J'_2 I F'_2 F'_{12} L' F' M'_F 
    \left | \hat{H}_{S}(1) \right | \beta_1 \beta_2 J_1 I F_1 J_2 I F_2 F_{12} L F M_F 
    \right \rangle \nonumber \\
  & = \mathcal{E} \delta_{\beta'_2 \beta_2} \delta_{J'_2 J_2} \delta_{F'_2 F_2} 
    \delta_{L' L} \delta_{M'_F M_F} 
    (-1)^{I+J'_1 +F'_1 + F'_{12} + F_1 +F_2 +F_{12}+L+F} 
    \sqrt{(2F'_1 +1)(2F_1 +1)(2F'_{12} +1)(2F_{12} +1)(2F+1)} \nonumber \\ 
  &	\times \sixj{J_1}{I}{F_1}{F'_1}{1}{J'_1}
    \sixj{F_1}{F_2}{F_{12}}{F'_{12}}{1}{F'_1} \sixj{F_{12}}{L}{F}{F'}{1}{F'_{12}} 
    C_{FM_F 10}^{F' M_F} \left \langle \beta'_1 J'_1 \left \| \hat{Q}_{1}(1) 
    \right \|\beta_1 J_1 \right \rangle .
  \label{eq:stark-fcp-1} 
\end{align}	
Due to the cylindrical symmetry about the $z$ axis, the total angular momentum projection $M_F$ is a good quantum number ($M'_F=M_F$), while its magnitude obeys the selection rule $F'=F$ or $F\pm 1$.

\subsubsection{Zeeman Hamiltonian}

When applying a homogeneous external magnetic field $\bm{B} = B \bm{u}_z$ along the $z$ direction, the Zeeman Hamiltonian can be written as $\hat{H}_Z = \hat{H}_Z(1) + \hat{H}_Z(2) = -(\mu_{10}(1)+\mu_{10}(2))B$, which gives in the uncoupled basis
\begin{align}
  & \left \langle \beta'_1 \beta'_2 J'_1 I F'_1 M'_{F_1} J'_2 I F'_2 M'_{F_2} L' M'_L 
    \left | \hat{H}_Z \right | 
    \beta_1 \beta_2 J_1 I F_1 M_{F_1} J_2 I F_2 M_{F_2} L M_L \right \rangle
  \nonumber \\
  & = \delta_{L' L} \delta_{M'_L M_L} \left[ \delta_{\beta'_2 \beta_2} 
    \delta_{J'_2 J_2} \delta_{F'_2 F_2} \delta_{M'_{F_2} M_{F_2}}
    \left \langle \beta'_1 J'_1 I F'_1 M'_{F_1} \left | \hat{H}_Z(1) 
    \right | \beta_1 J_1 I F_1 M_{F_1} \right \rangle \right. \notag \\
  & \left. \quad + \delta_{\beta'_1 \beta_1} \delta_{J'_1 J_1} 
    \delta_{F'_1 F_1} \delta_{M'_{F_1} M_{F_1}} 
    \left \langle \beta'_2 J'_2 I F'_2 M'_{F_2} \left | \hat{H}_Z(2) \right | 
    \beta_2 J_2 I F_2 M_{F_2} \right \rangle  \right] 
  \nonumber \\
  & = -B \delta_{L' L} \delta_{M'_L M_L} \left[ \delta_{\beta'_2 \beta_2} 
    \delta_{J'_2 J_2} \delta_{F'_2 F_2} \delta_{M'_{F_2} M_{F_2}}
    \left \langle \beta'_1 J'_1 I F'_1 M'_{F_1} \left | \hat{\mu}_{10}(1) 
    \right | \beta_1 J_1 I F_1 M_{F_1} \right \rangle \right. \notag \\
  & \left. \quad + \delta_{\beta'_1 \beta_1} \delta_{J'_1 J_1} 
    \delta_{F'_1 F_1} \delta_{M'_{F_1} M_{F_1}} 
    \left \langle \beta'_2 J'_2 I F'_2 M'_{F_2} \left | \hat{\mu}_{10}(2) 
    \right | \beta_2 J_2 I F_2 M_{F_2} \right \rangle  \right] .
\end{align}
Recalling that the matrix elements of $\hat{\mu}_{10}(i)$ are such that $\beta'_i = \beta_i$ and $J'_i=J_i$ (see Eq.~\eqref{eq:magmom}), and using the relations given in Appendix \ref{sec:cg}, we get to the fully-coupled basis expression
\begin{align}
  & \left \langle \beta'_1 \beta'_2 J'_1 I F'_1 J'_2 I F'_2 F'_{12} L' F' M'_F 
    \left | \hat{H}_Z \right | \beta_1 \beta_2 J_1 I F_1 J_2 I F_2 F_{12} L F M_F 
    \right \rangle \nonumber \\
  & = -\mu_{B} B \delta_{\beta'_1 \beta_1} \delta_{J'_1 J_1} \delta_{\beta'_2 \beta_2} 
    \delta_{J'_2 J_2} \delta_{L' L} \delta_{M'_F M_F} 
    \sqrt{(2F'_{12} +1)(2F_{12} +1)(2F+1)}
    \sixj{F_{12}}{L}{F}{F'}{1}{F'_{12}} C_{FM_F 10}^{F' M_F}
  \nonumber \\
  & \times \left[ (-1)^{I+J'_1 +F'_1 + F'_{12} + F_1 +F_2 +F_{12}+L+F} 
    \delta_{F'_2 F_2} g_{1} \sixj{J_1}{I}{F_1}{F'_1}{1}{J'_1} 
    \sixj{F_1}{F_2}{F_{12}}{F'_{12}}{1}{F'_1} \right. 
  \nonumber \\ 
  & \quad \left. \times \sqrt{J_1(J_1+1)(2J_1+1)(2F'_1 +1)(2F_1 +1)} \right. 
  \nonumber \\
  & \quad \left. + (-1)^{I+J'_2-F_1+L+F} 
    \delta_{F'_1 F_1} g_{2} \sixj{J_2}{I}{F_2}{F'_2}{1}{J'_2} 
    \sixj{F_2}{F_1}{F_{12}}{F'_{12}}{1}{F'_2} \right. 
  \nonumber \\ 
  & \quad \left. \times \sqrt{J_2(J_2+1)(2J_2+1)(2F'_2 +1)(2F_2 +1)} \right] .
  \label{eq:zeem}
\end{align}
Similarly to the Stark Hamiltonian $\hat{H}_S$, the quantum number $M_F$ is conserved while $F$ is not ($F'=F$ or $F\pm 1$).

\subsubsection{Long-range potential energy}

The matrix element of the space-fixed long-range operator $\hat{V}_{LR}(R)$ in the uncoupled basis can be expressed as \cite{lepers2018},
\begin{align}
  & \left \langle \beta'_1 \beta'_2 J'_1 I F'_1 M'_{F_1} J'_2 I F'_2 M'_{F_2} L' M'_L 
    \left | \hat{V}_{LR}(R) \right |
    \beta_1 \beta_2 J_1 I F_1 M_{F_1} J_2 I F_2 M_{F_2} L M_L \right \rangle 
  \nonumber \\
  & = \frac{X_0}{4\pi} \sum_{\ell_1 \ell_2 \ell} \delta_{\ell_1+\ell_2,\ell} 
	\frac{(-1)^{\ell_2}}{R^{\ell+1}} \sqrt{\frac{(2\ell)!}{(2\ell_1)!(2\ell_2)!}} 
	\sqrt{\frac{2L+1}{2L'+1}} C_{L0\ell 0}^{L'0} 
  \nonumber \\
  & \times \frac{\left \langle \beta'_1 J'_1 I F'_1 \left \| \hat{Q}_{\ell_1}(1) 
    \right \| \beta_1 J_1 I F_1 \right \rangle} {\sqrt{2F'_1+1}} 
    \frac{\left \langle \beta'_2 J'_2 I F'_2 \left \| \hat{Q}_{\ell_2}(2) 
    \right \| \beta_2 J_2 I F_2 \right \rangle} {\sqrt{2F'_2+1}} 
  \nonumber \\
  & \times \sum_{m m_1 m_2}(-1)^m C_{\ell_1 m_1 \ell_2 m_2}^{\ell m} 
    C_{L M_L \ell-m}^{L' M'_L} C_{F_1 M_{F_1} \ell_1 m_1}^{F'_1 M'_{F_1}}
    C_{F_2 M_{F_2} \ell_2 m_2}^{F'_2 M'_{F_2}}
  \label{eq:vlr-uncpl}
\end{align}
where $X_0=1/\epsilon_0$ or $\mu_0$ for electric and magnetic multipoles respectively, and $n!$ is the factorial of $n$. The positive integers $\ell_1$ and $\ell_2$ are the ranks of the atomic multipole moments. In this work, we deal with dipole $\ell_i=1$ and quadrupole moments $\ell_i=2$. The third index $\ell$ is the sum of $\ell_1$ and $\ell_2$; the Clebsch-Gordan coefficient of the second line of Eq.~\eqref{eq:vlr-uncpl} imposes that $L+\ell+L'$ is even. Recalling that $p_i$ is the parity under inversion of the electronic coordinates about each atomic nucleus, the electric multipole ranks are such that $p_i p'_i (-1)^{\ell_i} = 1$. In other words, the dipole moment changes the parity, while the quadrupole moment does not, as well known.

By using the relations in Appendix \ref{sec:cg} and the reduced matrix elements of Eqs.~\eqref{eq:qlm-we-1} and \eqref{eq:qlm-we-2}, we can derive the matrix element formula in the fully coupled basis,
\begin{align}
  & \left \langle \beta'_1 \beta'_2 J'_1 I F'_1 J'_2 I F'_2 F'_{12} L' F' M'_F 
    \left | \hat{V}_{LR}(R) \right | 
    \beta_1 \beta_2 J_1 I F_1 J_2 I F_2 F_{12} L F M_F \right \rangle 
  \nonumber \\
  & = \frac{X_0}{4\pi} \delta_{F' F} \delta_{M'_F M_F}
    \sum_{\ell_1 \ell_2 \ell} \delta_{\ell_1+\ell_2,\ell} 
    \frac{(-1)^{\ell_1 + J'_1 + J'_2 + 2I + F_1 + F_2 + F_{12}+L'+F}}{R^{\ell+1}}
    \sqrt{\frac{(2\ell)!}{(2\ell_1)!(2\ell_2)!}} C_{L0\ell 0}^{L'0}
  \nonumber \\
  & \times \sqrt{(2F'_1+1)(2F_1+1)(2F'_2+1)(2F_2+1)(2F'_{12}+1)(2F_{12}+1)
    (2L+1)(2\ell+1)} 
  \nonumber \\
  & \times \sixj{J_1}{I}{F_1}{F'_1}{\ell_1}{J'_1} 
    \sixj{J_2}{I}{F_2}{F'_2}{\ell_2}{J'_2} \sixj{F_{12}}{L}{F}{L'}{F'_{12}}{\ell} 
    \ninej{F'_{12}}{F'_1}{F'_2}{F_{12}}{F_1}{F_2}{\ell}{\ell_1}{\ell_2} 
    \left \langle \beta'_1 J'_1 \left \| \hat{Q}_{\ell_1}(1) 
    \right \|\beta_1 J_1 \right \rangle \left \langle \beta'_2 J'_2 
    \left \| \hat{Q}_{\ell_2}(2) \right \|\beta_2 J_2 \right \rangle
\end{align}	
where the last quantity between curly brackets is a Wigner 9-j symbol. In the fully-coupled basis, the long-range potential conserves the total angular momentum $F$ and its projection $M_F$.
%
\end{widetext}

\section{Results and discussions}
\label{sec:rd}

\begin{table*}
	\caption{Spectroscopic parameters used in this work, including energies, Land\'e $g$-factors, hyperfine constants, radiative lifetimes, and reduced transition dipole and quadrupole moments}
	\label{tab:para}
	\begin{ruledtabular}
		\begin{tabular}{rrrrrrrr}
			\multirow{2}{*}{Term} & \multirow{2}{*}{Level} & \multirow{2}{*}{Parity} & $E^\mathrm{a}$ & Land\'e & \multicolumn{2}{c}{Hyperfine constants} & Radiative  \\
			\cline{6-7}
			& &  & (cm$^{-1}$) & $g$-factor & $\phantom{A^{A^A}}A$ (MHz) & $\phantom{A^{A^A}}B$ (MHz) & lifetime (ns)  \\
			\hline                                                                   
			$4f^{11}({^4}I_{15/2}^{o})6s^{2}({^1}S_0)$ ${^4}I_{15/2}^{o^{\phantom{A}}}$ &  $|g \rangle$ &        odd        &           0 &          1.195$^\mathrm{a}$ &  800.583$^\mathrm{c}$ & -1688.0$^\mathrm{c}$ & - \\
			$4f^{11}({^4}I^{o})5d6s({^1}D)$ ${^4}I_{15/2}^{o}$  & $|a\rangle$ & odd &     24357.90     &   1.181$^\mathrm{b}$ &     840.3$^\mathrm{d}$  &             -1873.7$^\mathrm{d}$ & 3200$^\mathrm{b}$  \\ 
			$4f^{11}({^4}I_{15/2}^{o})6s6p({^1}P_1^{o})$ $(15/2,1)_{17/2}$  &  $|b \rangle$ &     even     & 24360.81 &     1.176$^\mathrm{b}$ &   654.9$^\mathrm{e}$ & -620.0$^\mathrm{e}$ & 4.9$^\mathrm{f}$  \vspace{5mm}\\
			\cline{1-2} 
			$\phantom{A^{A^A}}$
			reduced transition & absolute \\
			multipole moment & value$^\mathrm{b}$ (a.u.) \\
			\cline{1-2}
			$\phantom{A^{A^A}}$
			$|\langle a \| \hat{Q}_{1} \| b \rangle |$ & 2.56 \\
			$|\langle g \| \hat{Q}_{1} \| b \rangle |$ & 11.6 \\
			$|\langle g \| \hat{Q}_{2} \| a \rangle |$ & 35.3	
		\end{tabular}
	\end{ruledtabular}
	$^\mathrm{a}$ from NIST database \cite{NIST_ASD}, $^\mathrm{b}$ values calculated with Cowan code \cite{cowan1981} as in Ref.~\cite{li2017a} \\
	$^\mathrm{c}$ from Ref.~\cite{dankwort1974, wyart1978}, 
	$^\mathrm{d}$ from Ref.~\cite{Boutalib1992}, $^\mathrm{e}$ from Ref.~\cite{newman2011, miao2014}, $^\mathrm{f}$ from Ref.~\cite{den-hartog1999}
\end{table*}

Holmium has a single stable isotope, $^{165}$Ho, which is bosonic with a nuclear spin $I=7/2$. The electronic configuration and term of the ground level $|g\rangle$ are [Xe]$4f^{11}6s^{2} \,^{4}I_{15/2}^{o}$, with $J_g=15/2$. There exists a pair of quasi-degenerate levels, separated by 2.9 cm$^{-1}$: the odd-parity level $|a \rangle$ at 24357.90 cm$^{-1}$ with $J_{a}=15/2$ and the even-parity level $|b \rangle$ at 24360.81 cm$^{-1}$ with $J_{b}=17/2$.

The necessary spectroscopic data for the three levels are listed in Table~\ref{tab:para}, in particular the reduced transition multipole moments. The strong electric dipole-allowed transition between $|a\rangle$ and $|b\rangle$ comes from the $^{1}D-{}^{1}P^{o}$ character of the valence shells, while the transition between $|g\rangle$ and $|b\rangle$ comes from the $^{1}S-{}^{1}P^{o}$ character. Besides, there is a significant electric quadrupole-allowed transition between $|g\rangle$ and $|a\rangle$, due to $^{1}S-{}^{1}D$ character.  
These large transition multipole moments result in strong resonant interactions $\hat{V}_{dd}^{res}(R)$, $\hat{V}_{qq}^{res}(R)$ and $\hat{V}_{dq/qd}^{res}(R)$ in comparison with the direct ones $\hat{V}_{qq}^{dir}(R)$, $\hat{V}_{dq}^{dir}(R)$ and $\hat{V}_{\mu\mu}^{dir}(R)$ (see Table~\ref{tab:uncp}), which will be ignored in what follows. Indeed the direct interactions are proportional to the weak permanent quadrupole moment of $|g\rangle$, that we estimate on the order of 1 atomic unit \cite{lepers2016a}.

In 2014, sub-Doppler laser cooling and magneto-optical trapping of holmium was demonstrated by using the transition at 410.5 nm \cite{miao2014}, between the highest hyperfine levels of $|g\rangle$ and $|b\rangle$, of total angular momentum $F_g=11$ and $F_b=12$ respectively. Assuming ultracold spin-polarized atoms in the highest Zeeman sublevel $M_{F_g}=11$ and colliding in $s$-wave ($L=M_L=0$), we can deduce that a pair of colliding atoms possesses a total angular momentum projection $M_F=22$.
If we consider that the atom pair is submitted to a linearly-polarized photoassociation (PA) laser, red-detuned with respect to an atomic transition involving levels $|a\rangle$ or $|b\rangle$, the excited pair will also possess a total angular momentum projection $M_F=22$.

In this article, we model the excited pair of atoms, \textit{i.e.}~once a PA photon has been absorbed, submitted to collinear static electric and magnetic fields. Therefore, the total angular momentum projection $M_F$ is conserved by the Hamiltonian \eqref{eq:hamilt}, but not the total angular momentum $F$.
In order to perform our calculations, we set $F$ in a range $F_{min}=22 \leqslant F \leqslant F_{max}=27$, and $0 \leqslant L \leqslant L_{max}=4$ (except otherwise stated). This gives 752 potential-energy curves, in which 252 dissociate to the $|g\rangle+|a\rangle$ asymptotes, and 500 dissociate to the $|g\rangle+|b\rangle$ asymptote. In the symmetrized basis, the number of curves decreases down to 376, among which 126 dissociate to the $|g\rangle+|a\rangle$ asymptotes.

\begin{figure*}
	\begin{center}
		\includegraphics[width=0.328\textwidth]{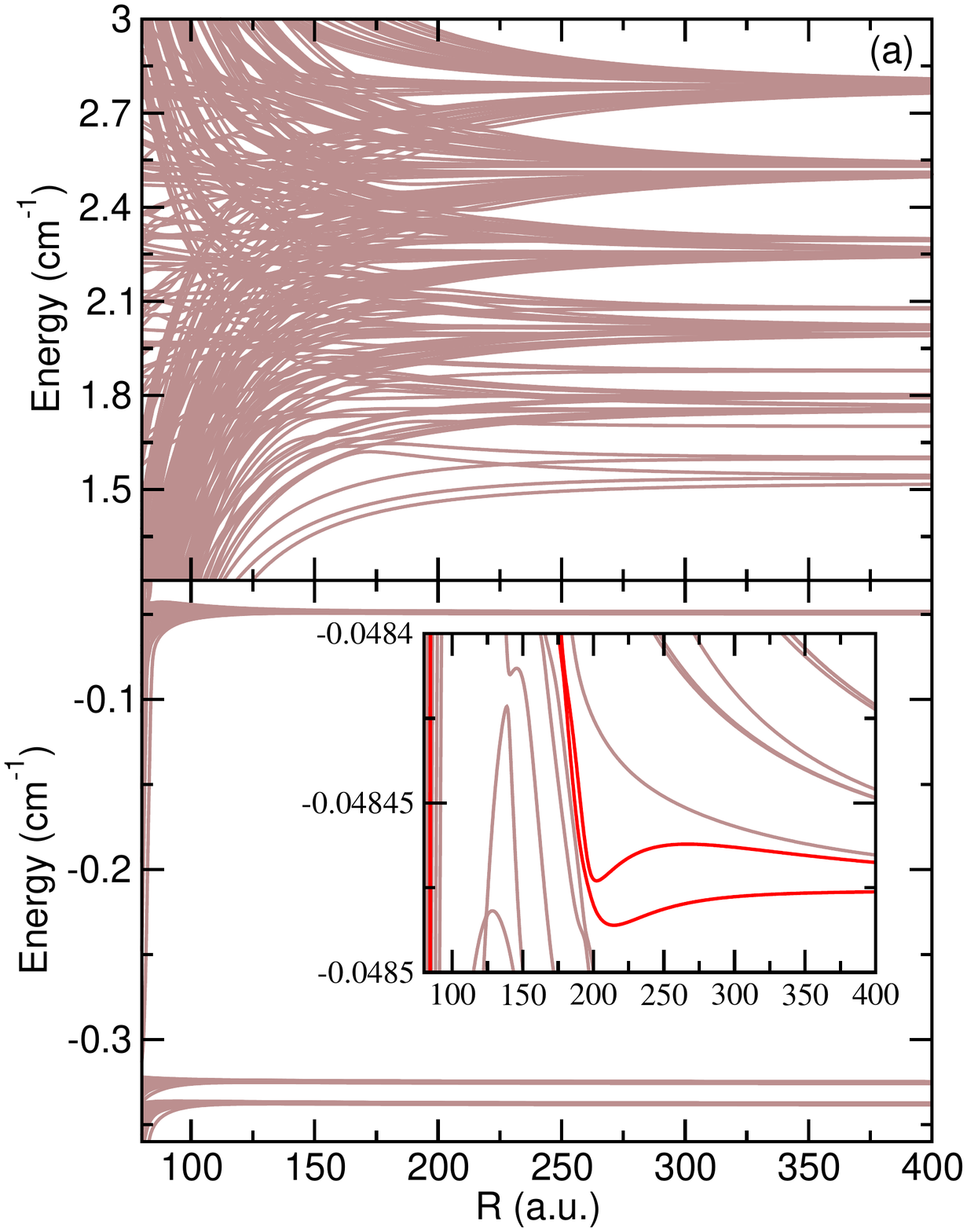}
		\includegraphics[width=0.328\textwidth]{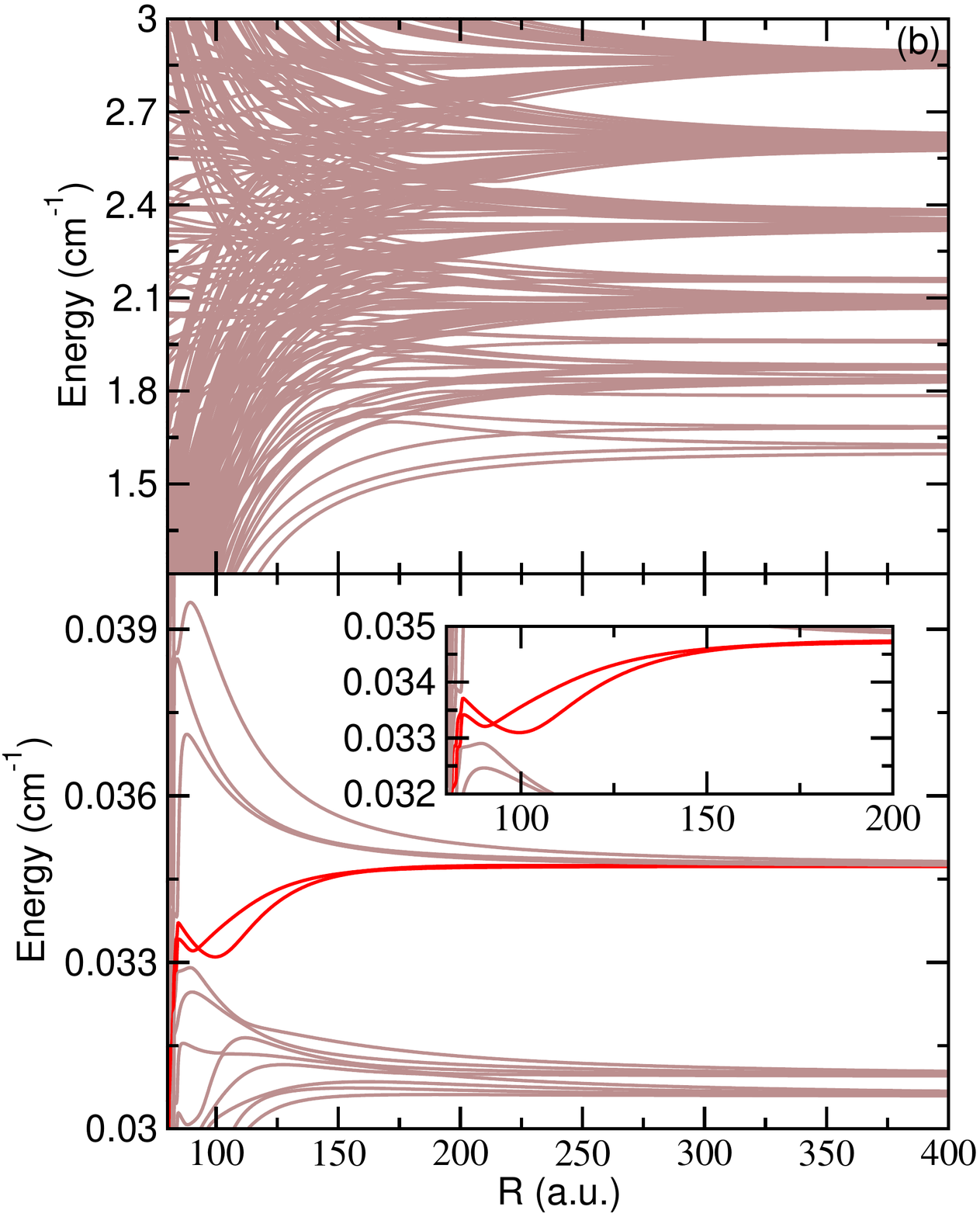}
		\includegraphics[width=0.328\textwidth]{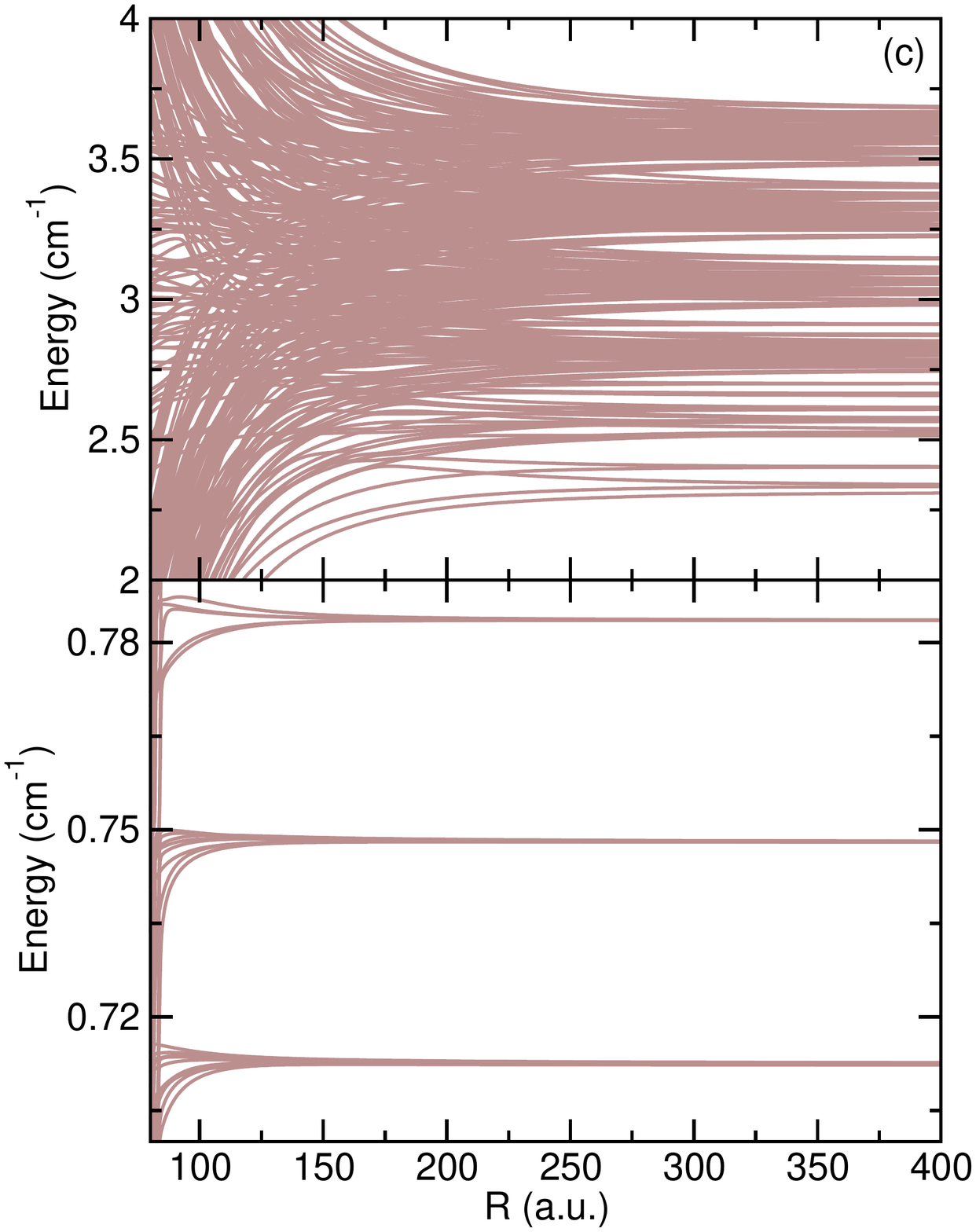}
	\end{center}
	\caption{Potential-energy curves for $\mathcal{E} = 5$~kV/cm and: (a) $B=0$~G, (b) $B=100$~G, and (c) $B=1000$~G, as functions of internuclear separation $R$. The upper panels correspond to curves converging towards the different hyperfine $|g\rangle+|b\rangle$ asymptotes, and the lower panels towards the highest levels of the $|g\rangle+|a\rangle$ asymptotes. The insets are zooms showing long-range potential wells. The angular momenta are $M_F=22$, $0\le L\le 4$, $22\le F\le 27$, which implies $7\le F_g\le 11$, $7\le F_a\le 11$ for the $|g\rangle+|a\rangle$, and $6\le F_g\le 11$, $7\le F_b\le 12$ for the $|g\rangle+|b\rangle$ asymptote.
		\label{fig:L4E5}  
	}
\end{figure*}

From our long-range model, calculated potential-energy curves for excited atom pairs are presented in Figs.~\ref{fig:L4E5}(a)--(c) for $\mathcal{E} = 5$~kV/cm, $B=0$, 100 and 1000~G, which are typical experimental values. The lower and upper panels show the curves dissociating towards the $|g\rangle+|a\rangle$ and $|g\rangle+|b\rangle$ asymptotes respectively. The zero energy has been set to the average of the field-free $|a\rangle$ and $|b\rangle$ level energies, though it is not visible in Fig.~\ref{fig:L4E5} due to the break in the energy scale. Figures~\ref{fig:L4E5}(a)--(c) display all the curves converging to $|g\rangle+|b\rangle$ with $M_F=22$, $0 \le L \le 4$ and $22 \le F \le 27$, which explains why the curves converge to the 6 highest hyperfine levels of $|g\rangle$ ($6\le F_g\le 11$) and $|b\rangle$ ($7\le F_b\le 12$) since $18\le F_{12}\le 23$. At this range of internuclear distances, the curves are strongly mixed due to the $R^{-3}$-dependent resonant dipole-dipole interaction. By contrast, the curves close to the $|g\rangle+|a\rangle$ asymptote are less mixed, because of the shorter-range $R^{-5}$-dependent resonant quadrupole-quadrupole interaction. Figures \ref{fig:L4E5}(a)--(c) only contain the highest curves converging to $|g\rangle+|a\rangle$, because if we were showing all of them, we would only see flat curves. Finally the insets comprise some selected curves containing long-range potential wells.



We can see that those wells become deeper with increasing magnetic field, and that their minimum is shifted to smaller internuclear distances. As shown on Fig.~\ref{fig:L4E5}, they are coupled to the attractive curves coming from the $|g\rangle + |b\rangle$ asymptote, inducing a potential barrier toward small distances. The possibility for the atoms to cross this barrier may reduce the lifetime of vibrational levels, since these atoms are very likely to experience inelastic collisions in the small-$R$ region. Predissociation due to non-adiabatic couplings between potential wells and lower repulsive curves may also limit the lifetime, as well as spontaneous emission. For atom pairs close to $|g\rangle + |a\rangle$ asymptotes, due to the low electric fields in this study, we can assume that the radiative lifetime is close to the one of level $|a\rangle$ (see Table~\ref{tab:para}). However increasing the field amplitude increases the mixing with level $|b\rangle$, and so decreases the radiative lifetime since $\tau_b$ is 650 times smaller than $\tau_a$.

\begin{figure}
	\begin{center}
		\includegraphics[width=0.5\textwidth]{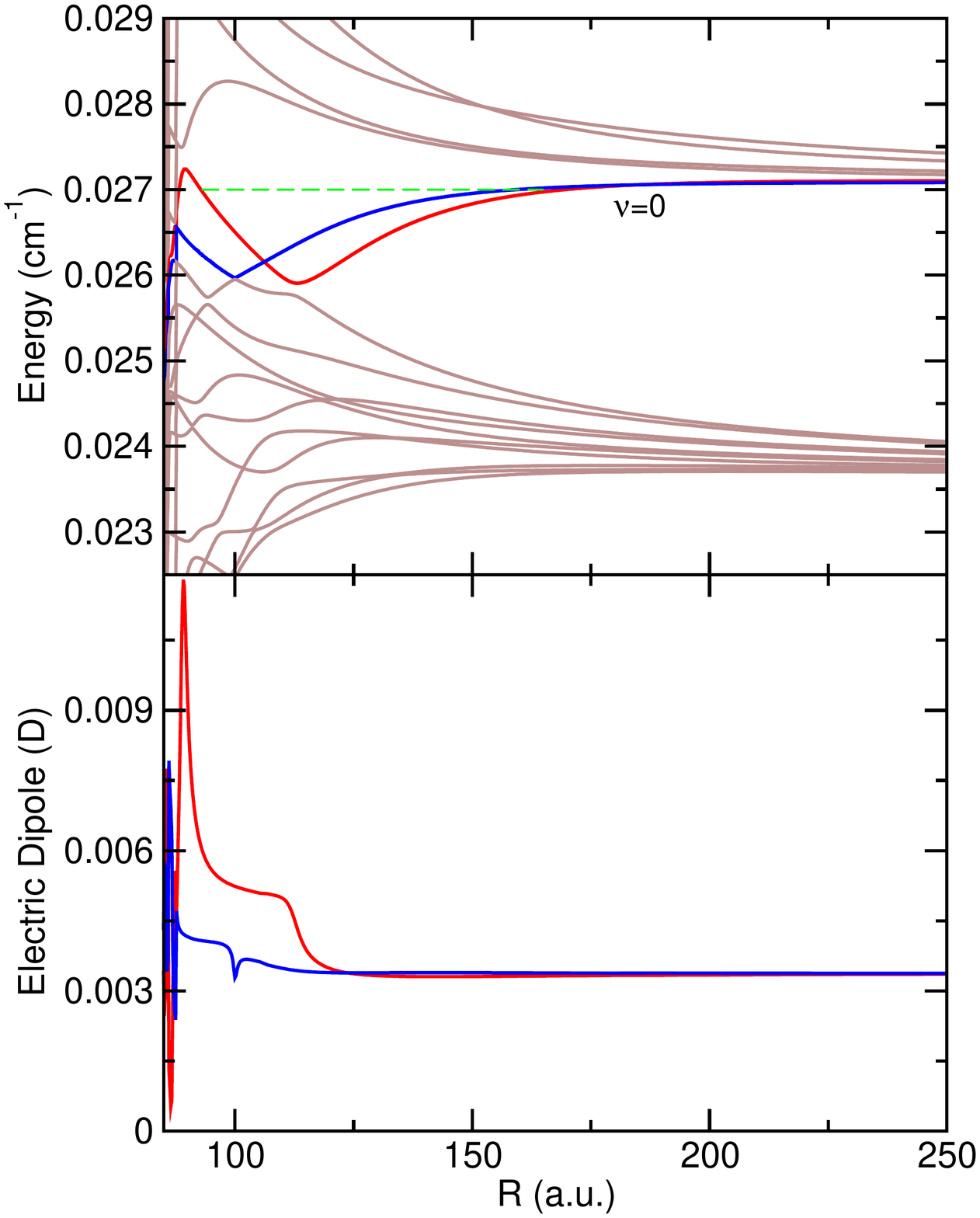}
	\end{center}
	\caption{Upper panel: zoom on selected levels of $|g\rangle+|a\rangle$ asymptote for $\mathcal{E} = 1$~kVv/cm and $B=90$~G ($L_{max}=5$). The dashed line represents a vibrational level supported by the heavy red potential well. Lower panel: induced electric dipole moment associated with the two heavy curves of the upper panel.
		\label{fig:ge1}  
	}
\end{figure}

In consequence, there is a compromise to find in terms of field amplitudes. A larger magnetic field deepens the potential wells, hence favoring the existence of vibrational levels, but it can reduce their lifetimes because of an easier tunneling toward small internuclear distances. Also, a larger electric field increases the induced dipole moment through a larger coupling between levels $|a\rangle$ and $|b\rangle$, but it reduces the radiative lifetime. A good compromise can be found with $\mathcal{E} = 1$~kV/cm and $B=90$~G, as shown in Fig.~\ref{fig:ge1}. The red curve is a $10^{-3}$-cm$^{-1}$-deep potential well correlated to the asymptote ($F_g=M_{F_g}=11$, $F_a=M_{F_a}=11$, $F_{12}=22$), and which supports a vibrational level. It was computed with the mapped-Fourier-grid-Hamiltonian (MFGH) method \cite{kokoouline1999, kokoouline2000} for the red potential-energy curve separately, which explains why we could not calculate the rate of predissociation toward lower curves.

The lower panel of Fig.~\ref{fig:ge1} shows the $R$-dependent induced electric dipole moment along $z$ for the highlighted curves of the upper panel, which are equal to a few thousandths of debye. They are $R$-independent for distances above 125~a.u., meaning that the mixing of levels $|a\rangle$ and $|b\rangle$ is very close to that with separated atoms ($R\to\infty$). The strong $R$-variation in the region of the well minimum indicates strong coupling between molecular states. Because the highlighted curves are characterized by the largest possible values of $F_{12}$, $M_{F_g}$ and $M_{F_a}$, they possess a strong magnetic moment (in absolute value), close to the extremal value $-(g_g J_g + g_a J_a) = -17.8~\mu_B$ corresponding to the sum of two separated atoms.

We can associate with this magnetic moment a characteristic dipolar length $a_d=md^2$  \cite{frisch2015}, $m$ being the mass of a Ho$_2$ molecule and $d$ the dipole moment (all quantities are expressed in atomic units). Taking $d=-17~\mu_B = -0.062$~a.u., we obtain $a_d=2300$~a.u., which is larger than the value of 1150~a.u.~corresponding to Er$_2$ Feshbach molecules \cite{frisch2015}. We can also estimate the dipolar length associated with the induced electric dipole moment $d=0.003~\mathrm{D} =0.0012$~a.u., which gives 0.84~a.u.. These two characteristic lengths open the possibility to observe dipolar effects with the magnetic dipoles, and manipulate the molecules with an external electric field.

\begin{figure*}
	\begin{center}
		\includegraphics[width=0.4\textwidth]{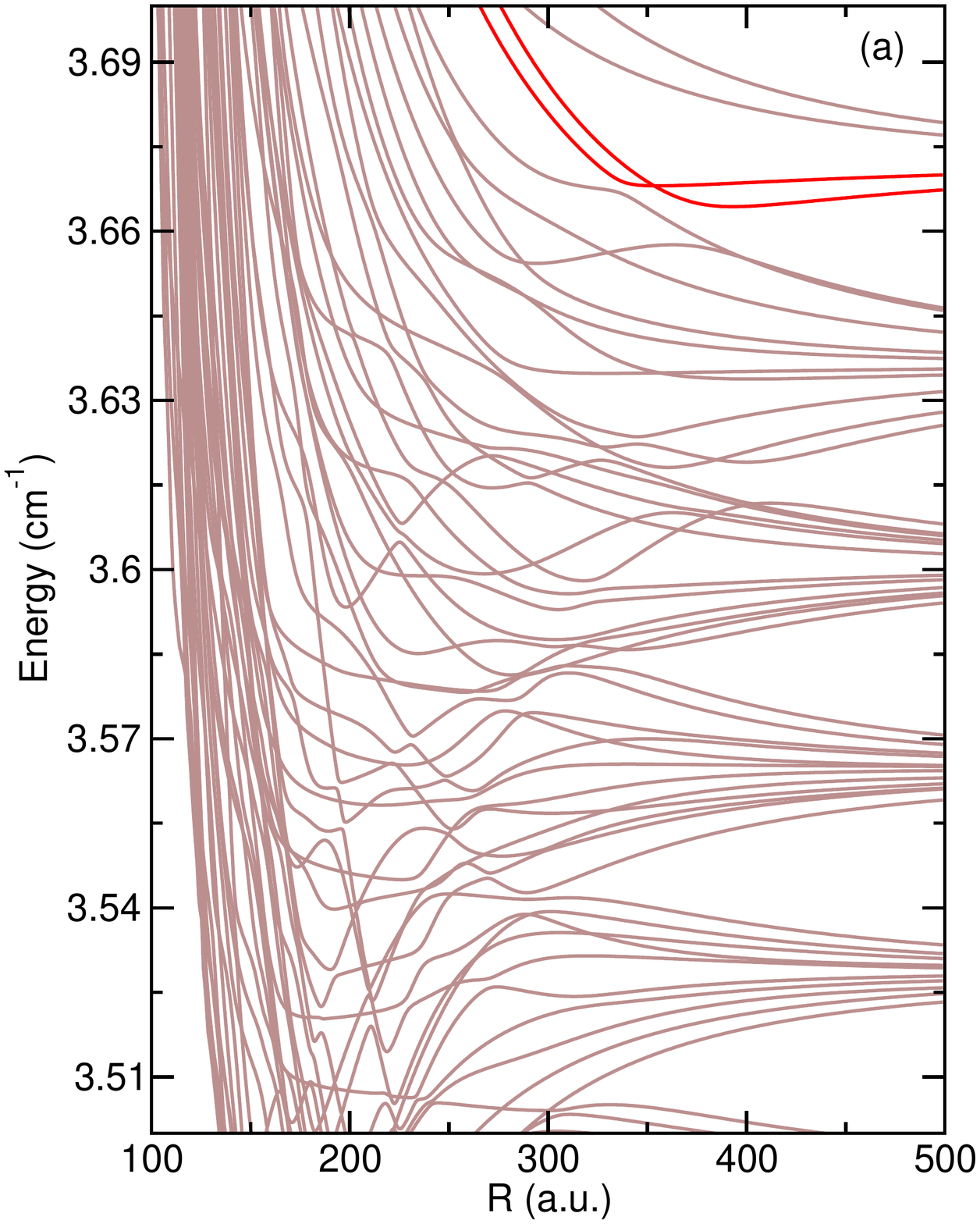}
		\includegraphics[width=0.4\textwidth]{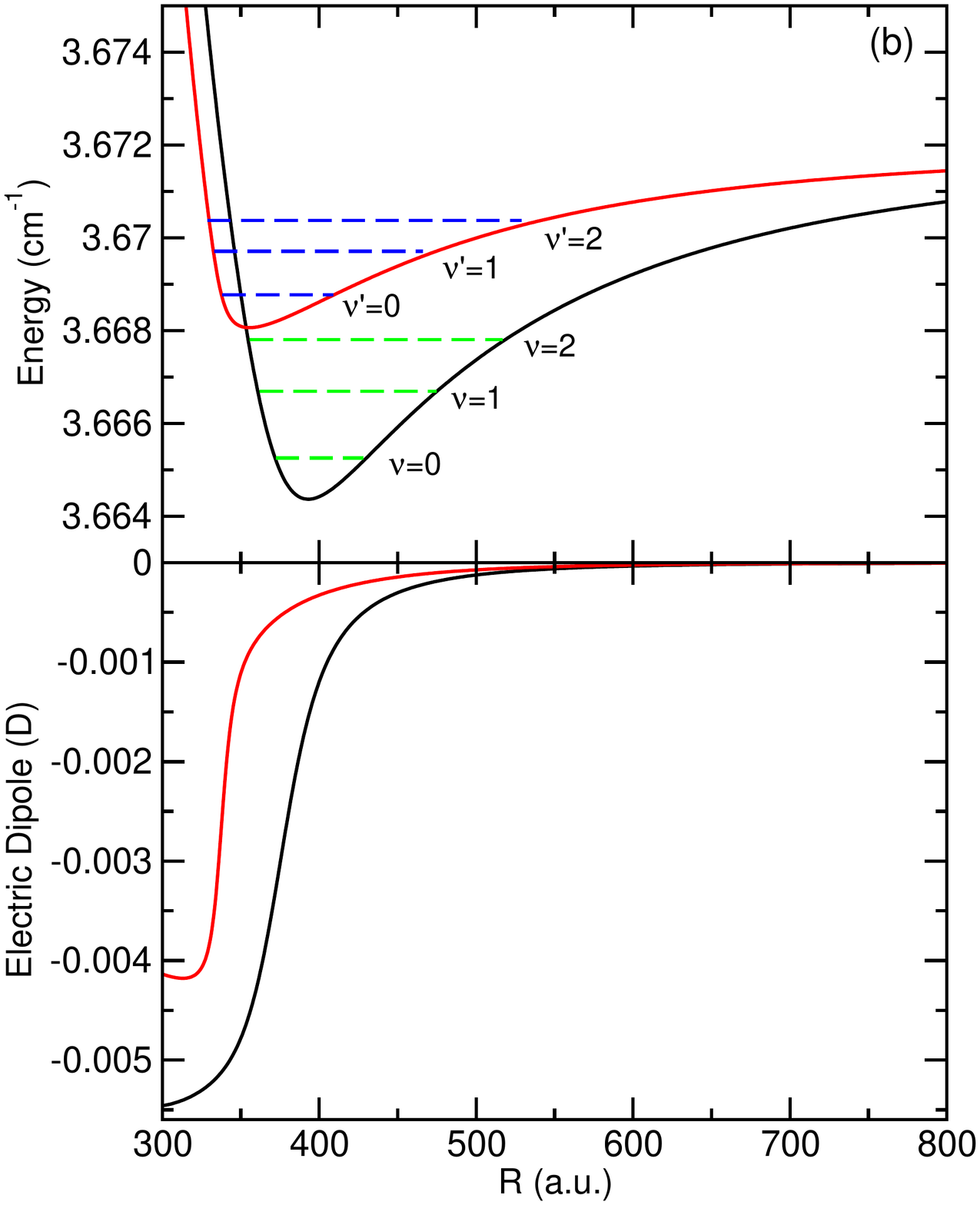}
	\end{center}
	\caption{(a) The highest potential energy curves obtained for $\mathcal{E} = 5$~kV/cm and $B=1000$~G (zoom of Fig.~\ref{fig:L4E5}(c)) as a function of internuclear separation $R$. Two long-range potential wells are highlighted in red. Upper panel of (b): those two wells with their computed three lowest vibrational levels. Lower panel: their $R$-dependent induced electric dipole moment.
		\label{fig:ge2}  
	}
\end{figure*}

Figure \ref{fig:ge2} shows that long-range wells with vibrational levels can also exist close to the $|g\rangle + |b\rangle$ asymptotes (actually $F_g=M_{F_g}=11$, $F_b=M_{F_b}=12$, $F_{12}=23$). These wells, which are deeper and longer-range than those of Fig.~\ref{fig:ge1}, support more vibrational levels (only the three lowest are shown for each well). Besides, these wells do not possess potential barriers allowing tunneling toward smaller distances. But, being close to the $|g\rangle + |b\rangle$ asymptotes, their radiative lifetime is approximately $\tau_b = 4.9$~ns. Their induced dipole moments equal a few thousandths of debye as in Fig.~\ref{fig:ge1}, but here they tend to 0 in the separated-atom limit. This is because the sublevel $F_b=M_{F_b}=12$ is insensitive to the electric field, as it is cannot be mixed with any sublevel $(F_a,M_{F_a})$ since $M_{F_a}\le 11$ (see Ref.~\cite{lepers2018a} for a detailed discussion).

\section{Conclusion}
\label{sec:conclu}

We have calculated the long-range potential energy between two identical lanthanide atoms, one in the ground level and one in a superposition of nearly degenerate excited levels coupled by an external electric field. This situation gives rise to various direct and resonant interactions between atomic multipoles. Our formalism that includes the atomic hyerfine structure is presently applied to holmium, but it can also be applied to other lanthanide atoms with quasi-degenerate energy levels \cite{lepers2018a}, or to Rydberg atoms with a large angular momentum.

In the case of holmium, our calculations predict the existence of long-range potential wells that are likely to support vibrational levels, accessible by photoassociation from the ground level. Their strong magnetic moments make them interesting alternatives to the Feshbach Er$_2$ molecules formed by magneto-association \cite{frisch2015}, with in addition to Er$_2$, an induced electric dipole moment that opens the possibility to prepare and manipulate the molecules with an external electric field. Furthermore, the large number of repulsive curves opens the possibility of optical shielding, in order to control the collisions between ground-level holmium atoms. Another possibility is to bring a vibrational level very close to the dissociation limit so that one can tune the scattering length of such atoms \cite{lassabliere2018}, as it was proposed for molecular collisions in a microwave field \cite{lassabliere2018, karman2018}.

\section*{Acknowledgements}

The authors acknowledge support from {}``DIM Nano-K'' under the project {}``InterDy'', and from {}``Agence Nationale de la Recherche'' (ANR) under the project {}``COPOMOL'' (contract ANR-13-IS04-0004-01).

\appendix

\section{Fully-coupled symmetrized basis}
\label{sec:cplbas}

\subsection{Construction of the fully-coupled basis}

This appendix sketches the process of the construction of the fully coupled basis. For a single atom, the electronic angular momentum $\hat{\bm{J}_i}$ is coupled with the nuclear-spin angular momentum $\hat{\bm{I}}$ to form $\hat{\bm{F}_i} = \hat{\bm{J}_i} + \hat{\bm{I}}$, $i = 1, 2$, so that the state of each atom can be expressed as, 
\begin{equation}
  |\beta_i J_iIF_iM_{F_i} \rangle = \sum_{M_{J_i}M_{I_i}}
    C_{J_iM_{J_i}IM_{I_i}}^{F_iM_{F_i}} |\beta_i J_iM_{J_i}\rangle |IM_{I_i}\rangle
\end{equation}
Then, the total angular momenta of two atoms $\hat{\bm{F}}_i$ are coupled together to form $\hat{\bm{F}}_{12} = \hat{\bm{F}}_1 + \hat{\bm{F}}_2$, to give
\begin{equation}
  |F_1F_2F_{12}M_{F_{12}}\rangle = \sum_{M_{F_1}M_{F_2}}
    C_{F_1M_{F_1}F_2M_{F_2}}^{F_{12}M_{F_{12}}} |F_1M_{F_1}\rangle 
    |F_2M_{F_2}\rangle
\end{equation}
which is subsequently coupled to the rotational angular momentum $\hat{\bm{L}}$ to form the total angular momentum of the complex $\hat{\bm{F}} = \hat{\bm{F}}_{12} + \hat{\bm{L}}$, with projection $M_F$ along the space-fixed $z$ axis,
\begin{align}
  |F_{12}LFM_F\rangle = \sum_{M_{F_{12}}M_L} 
    C_{F_{12}M_{F_{12}}LM_L}^{FM_F} |F_{12} M_{F_{12}}\rangle |L M_L\rangle .
\end{align}
Finally the fully coupled basis can be derived from the uncoupled one,
\begin{align}
  & |\beta_1\beta_2J_1IF_1J_2IF_2F_{12}LFM_F \rangle \nonumber \\
  & = \sum_{\substack{M_{F_{12}}M_LM_{F_1}M_{F_2}\\M_{J_1}M_{I_1}M_{J_2}M_{I_2}}} 
    C_{F_{12}M_{F_{12}}LM_L}^{FM_F} C_{F_1M_{F_1}F_2M_{F_2}}^{F_{12}M_{F_{12}}} 
  \nonumber \\
  & \times C_{J_1M_{J_1}IM_{I_1}}^{F_1M_{F_1}} C_{J_2M_{J_2}IM_{I_2}}^{F_2M_{F_2}}
  \nonumber \\
  & \times |\beta_1J_1M_{J_1}IM_{I_1}\rangle
    |\beta_2J_2M_{J_2}IM_{I_2}\rangle |LM_L\rangle .
  \label{eq:cplbas}
\end{align}

\subsection{Basis symmetrization}

We denote $\bm{r}_i$ the coordinates of all the electrons inside atom $i$, and $\bm{R}$ the vector joining atom 1 and atom 2. Firstly, we apply the inversion operator $\hat{E}^*: \bm{r}_1 \rightarrow -\bm{r}_1 ; \bm{r}_2 \rightarrow -\bm{r}_2 ; \bm{R} \rightarrow -\bm{R}$ to the fully-coupled state constructed above,
\begin{align}
  & \hat{E}^*|\beta_1\beta_2J_1IF_1J_2IF_2F_{12}LFM_F \rangle
  \nonumber \\ 
  & = p_1p_2(-1)^L |\beta_1\beta_2J_1IF_1J_2IF_2F_{12}LFM_F \rangle ,
\end{align}
where $p_i$ stand for the electronic parity of individual atoms. The basis functions are thus divided into even and odd functions if $p_1p_2(-1)^L = \pm 1$, both cases being allowed.

Now we consider the operator that interchanges atoms 1 and 2, $\hat{P}_{12}: \bm{r}_1 \rightarrow \bm{r}_2 ; \bm{r}_2 \rightarrow \bm{r}_1 ; \bm{R} \rightarrow -\bm{R}$, which gives in the uncoupled basis
\begin{align}
  & \hat{P}_{12} |\beta_1J_1M_{J_1}IM_{I_1}\rangle
    |\beta_2J_2M_{J_2}IM_{I_2}\rangle |LM_L\rangle
  \nonumber \\
  & = (-1)^L |\beta_2J_2M_{J_2}IM_{I_2}\rangle
    |\beta_1J_1M_{J_1}IM_{I_1}\rangle |LM_L\rangle .
\end{align}
When transforming this equation in the fully-coupled basis, one needs to take care of the step leading to $\hat{\bm{F}}_{12}$, that is
\begin{align}
  & \hat{P}_{12} |F_1F_2F_{12}M_{F_{12}}\rangle
  \nonumber \\
  = & \hat{P}_{12} \sum_{M_{F_1}M_{F_2}} C_{F_1M_{F_1}F_2M_{F_2}}^{F_{12}M_{F_{12}}}
    |F_1M_{F_1}\rangle |F_2M_{F_2}\rangle
  \nonumber \\
  = & \sum_{M_{F_1}M_{F_2}} C_{F_1M_{F_1}F_2M_{F_2}}^{F_{12}M_{F_{12}}}
    \hat{P}_{12} |F_1M_{F_1}\rangle |F_2M_{F_2}\rangle 
  \nonumber \\
  = & \sum_{M_{F_1}M_{F_2}} C_{F_1M_{F_1}F_2M_{F_2}}^{F_{12}M_{F_{12}}}
    |F_2M_{F_2}\rangle |F_1M_{F_1}\rangle
  \nonumber \\
  = & (-1)^{F_1+F_2-F_{12}} \sum_{M_{F_1}M_{F_2}}
    C_{F_2M_{F_2}F_1M_{F_1}}^{F_{12}M_{F_{12}}}
    |F_2M_{F_2}\rangle |F_1M_{F_1}\rangle
  \nonumber \\
  = & (-1)^{F_1+F_2-F_{12}} |F_2F_1F_{12}M_{F_{12}}\rangle ,
\end{align}
where we used Eq.~\eqref{eq:cg-perm}. Finally, we obtain
\begin{align}
  & \hat{P}_{12} |\beta_1\beta_2J_1IF_1J_2IF_2F_{12}LFM_F \rangle
  \nonumber \\ 
  & = (-1)^{F_1+F_2-F_{12}+L} |\beta_2\beta_1J_2IF_2J_1IF_1F_{12}LFM_F\rangle ,
\end{align}
and so the symmetrized basis functions can be constructed as (see Eq.~\eqref{eq:sym-basis})
\begin{align}
  & |\beta_1\beta_2J_1IF_1J_2IF_2F_{12}LFM_F;\eta \rangle
  \nonumber \\
  = & \frac{1}{\sqrt{2(1+\delta_{\beta_1\beta_2}\delta_{J_1J_2}\delta_{F_1F_2})}}
  \nonumber \\
  \times & \left \{ |\beta_1\beta_2J_1IF_1J_2IF_2F_{12}LFM_{F} \rangle \right. 
  \nonumber \\
  & \left. + \eta (-1)^{F_1+F_2-F_{12}+L}
    |\beta_2\beta_1J_2IF_2J_1IF_1F_{12}LFM_F \rangle \right \}
\end{align}
where the symmetry of the basis functions with respect to the permutation of identical atoms is given by index $\eta$, which is equal to $+1$ for identical bosons and $-1$ for identical fermions \cite{tiesinga2005, hutson2006, quemener2011b}. There is only one possible value of $\eta$ for a given isotope.

In the special case $M_F=0$, the potential energy curves are divided into even $(\varepsilon=1)$ and odd $(\varepsilon=-1)$ ones with respect to the reflection $\sigma_{xz}$ about the space-fixed $xz$ plane. Because this reflection can be decomposed in an inversion followed by the rotation of $\pi$ radians around the $y$ axis. Since the latter transforms the basis function $|FM_F\rangle$ into $(-1)^{F-M_F} |F,-M_F\rangle$ \cite{varshalovich1988}, the even or odd character of a given basis function for $M_F=0$ is given by $\varepsilon = p_1p_2 (-1)^{L+F}$. The Clebsch-Gordan coefficients of the Stark and Zeeman Hamiltonians (see Eqs.~\eqref{eq:stark-fcp-2}, \eqref{eq:stark-fcp-1} and \eqref{eq:zeem}) impose $F'-F=\pm 1$ when $M_F=0$. Therefore the Stark Hamiltonian, which changes the total parity, conserves $\varepsilon$, whereas the Zeeman Hamiltonian, which conserves the total parity, changes $\varepsilon$.

\section{Relations involving Clebsch-Gorden coefficients used in this paper}
\label{sec:cg}

The relationships of this appendix are extracted from Ref.~\cite{varshalovich1988}, chapter 8. The permutation of lower indexes yields
\begin{equation}
  C_{b\beta a\alpha}^{c\gamma} = (-1)^{a+b-c}
    C_{a\alpha b\beta}^{c\gamma}
  \label{eq:cg-perm}
\end{equation}
Furthermore we list several important sums involving Clebsch-Gorden coefficients,
%
\begin{align}
& \sum_\alpha(-1)^{a-\alpha}C_{a\alpha a-\alpha }^{c0}=\sqrt{2a+1}\delta_{c,0} 
\\
& \sum_{\alpha \beta }C_{a\alpha b\beta }^{c\gamma }C_{a\alpha b\beta }^{{c}'\gamma '}=\delta_{c,c'}\delta_{\gamma,\gamma '}
\end{align}
\begin{align}
  & \sum_{\alpha\beta\delta} C_{a\alpha b\beta}^{c\gamma}
    C_{d\delta b\beta}^{e\varepsilon} C_{a\alpha f\varphi}^{d\delta}
  \nonumber \\
  & = (-1)^{b+c+d+f} \sqrt{(2c+1)(2d+1)} \begin{Bmatrix}
		a & b & c \\ 
		e & f & d
	\end{Bmatrix} C_{c\gamma f\varphi}^{e\varepsilon} 
\end{align}
\begin{align}
  & \sum_{\beta \gamma \varepsilon \varphi }C_{b\beta c\gamma }^{a\alpha }C_{ e\varepsilon f\varphi }^{ d\delta }C_{e\varepsilon g\eta}^{b\beta}C_{f\varphi j\mu }^{c\gamma } \nonumber \\
  & = \sum_{k\kappa }\sqrt{(2b+1)(2c+1)(2d+1)(2k+1)} \nonumber \\
  & \times C_{g\eta j\mu }^{k\kappa }C_{d\delta k\kappa }^{a\alpha }\begin{Bmatrix}
		a & b & c\\ 
		d & e & f\\ 
		k & g & j
    \end{Bmatrix}	
\end{align}
a explicit form of Clebsch-Gordan coefficient with special arguments,
\begin{equation}
C_{a \alpha 00}^{c \gamma} = \delta_{a, \alpha} \delta_{c, \gamma}
\end{equation}
A relation for a Wigner 9-j symbol with a zero argument reducing to 6-j symbol,
\begin{align}
& \ninej{a}{b}{c}{d}{e}{f}{g}{h}{0} =\ninej{f}{e}{d}{c}{b}{a}{0}{h}{g} \nonumber \\
& = \delta_{c,f} \delta_{g,h} \frac{(-1)^{b+c+d+g}}{\sqrt{(2c+1)(2g+1)}} \sixj{a}{b}{c}{e}{d}{g} .
\end{align}
%


%

\end{document}